\documentstyle[12pt]{article}
\font\titolo=cmbx12 scaled\magstep2

\font\tsnorm=cmr12

\font\tscorsp=cmti10
\def\CQG{ Class. Quant. Grav. }
\def\NCA{Nuovo Cimento }
\def\NPB{Nucl. Phys. }
\def\PLB{ Phys. Lett.  }  
\def\PRL{ Phys. Rev. Lett. }
\def\PRD{Phys. Rev.  } 
 
\def\MPLA{ Mod. Phys. Lett. } 
\def\CMP{ Comm. Math. Phys. }
\def\AP{ Ann. Phys. }
\def\z{Z\kern -4.6pt Z}

\def\xp{x^+}\def\xm{x^-}
\def\l{\lambda}
\def\lq{\l^2}

\def\ha{{1\over 2}}
\def\a{\alpha}
\def\b{\beta}

\def\d{\delta}
\def\f{\phi}
\def\g{\gamma}
\def\i{\iota}

\def\o{\omega}
\def\p{\pi}

\def\s{\sigma}

\def\y{\eta}
\def\z{\zeta}

\def\L{\Lambda}

\def\P{\Pi}

\def\lie{{\cal L}}
\def\de{\partial}
\def\na{\nabla}

\def\inf{\infty}

\def\mo{{-1}}
\def\ha{{1\over 2}}

\def\gmn{g_{\m\n}}

\def\ds{ds^2=}\def\sg{\sqrt{-g}}

\def\xk{\x^{(k)}}
\def\as{asymptotic symmetries }              
\def\eom{equations of motion }
\def\pb{Poisson brackets }

\def\na{\nabla}
\def\sg{\sqrt{-g}}

\def\d{\delta}
\def\e{\eta}
\def\eo{\e_0}
\def\m{\mu}
\def\n{\nu}

\def\gmn{g_{\m\n}}

\def\ord#1{o\left(#1\right)}
\def\xo{{\x^\perp}}
\def\xp{{\x^\parallel}}
\def\oo{{\o^\perp}}
\def\op{{\o^\parallel}}
\def\ul{{1\over\l}}
\def\Ht{{\cal H}}
\def\Hx{{\cal H}_x}
\def\Pe{\P_\y}
\def\Ps{\P_\s}
\def\xn{x^{+}}
\def\xm{x^{-}}

\def\s{\sigma}
\def\x{\chi}
\def\ds{ds^2=}

\def\la{\l^2}
\def\i{\infty}
\def\be{\begin{equation}}
\def\ee{\end{equation}}
\def\bea{\begin{eqnarray}}
\def\eea{\end{eqnarray}}
\def\bc{\begin{displaymath}}
\def\ec{\end{displaymath}}
\def\lb{\label}
\def\ads{Anti-de Sitter }
\def\adsd{$\rm AdS_{2}$ }
\def\adst{$\rm AdS_{3}$ }
\def\adsdp{$\rm AdS_{2}^{+}$ } 
\def\adsdm{$\rm AdS_{2}^{-}$ }
\def\adsdz{$\rm AdS_{2}^{0}$ }
\begin{document}
\pagestyle{empty}
\null
\vskip 5truemm
\begin{flushright}
INFNCA-TH9901\\
\end{flushright}
\vskip 15truemm
\begin{center}
{\titolo ASYMPTOTIC SYMMETRIES OF AdS$_{2}$ AND       }
\end{center}
\begin{center}
\titolo{ CONFORMAL GROUP IN d=1}
\end{center}
\vskip 15truemm
\begin{center}
{\tsnorm Mariano Cadoni$^{a,c,*}$ and Salvatore Mignemi$^{b,c,**}$}
\end{center}
\begin{center}
{$^a$\tscorsp Dipartimento di Fisica,  
Universit\`a  di Cagliari,}
\end{center}
\begin{center}
{\tscorsp Cittadella Universitaria, 09042, Monserrato, Italy.}
\end{center}
\begin{center}
{\tscorsp $^b$  Dipartimento di Matematica, Universit\`a  di Cagliari,}
\end{center}
\begin{center}
{\tscorsp viale Merello 92, 09123, Cagliari, Italy.}
\end{center}
\begin{center}
{\tscorsp $^c$  INFN, Sezione di Cagliari.}
\end{center}
\vskip 19truemm
\begin{abstract}
\noindent
We present a detailed discussion of the asymptotic symmetries of 
Anti-de Sitter space in two dimensions and their relationship with the 
conformal group in one dimension. We use this relationship to give
a microscopical derivation of the entropy of 2d black holes that 
have asymptotically Anti-de Sitter behaviour. The implications of our 
results for the conjectured \adsd/CFT$_{1}$ duality are also discussed.
\end{abstract}
\vfill
\hrule
\begin{flushleft}
{$^*$E-Mail: CADONI@CA.INFN.IT\hfill}
\end{flushleft}
\begin{flushleft}
{$^{**}$E-Mail: MIGNEMI@CA.INFN.IT\hfill}
\end{flushleft}
\vfill
\eject
\pagenumbering{arabic}
\pagestyle{plain}
\section{Introduction}
\paragraph{}

An important realization of the holographic principle \cite{Su}, stating that a 
bulk theory with gravity is equivalent to a boundary theory without 
gravity, is the Anti-de Sitter (AdS)/ conformal field theory (CFT) correspondence \cite{Wm}.
According to it, supergravity  on $d$-dimensional AdS space should be 
dual to a conformal field theory living on its  $d-1$-dimensional boundary.
For $d>4$ the AdS/CFT correspondence has been used   to gain
informations about  the nonperturbative regime of Yang Mills  theories,
i.e. to learn about field theory from gravity.
For $d<4$ the conformal symmetry is infinite dimensional,
so that one expects the opposite to be true, i.e. one should learn 
about gravity from field theory. 

A nice example of how this could work is represented by the $d=3$ case.  
It is well known  since the work of Brown and Henneaux \cite{Bh} that the 
asymptotic symmetry group of AdS$_{3}$ is the conformal group in two 
dimensions. This fact was the starting point of the investigations of
Strominger
who used two-dimensional (2d) conformal field theories results 
to understand black hole
physics.  He calculated  the entropy  of the 
three-dimensional (3d) Ba\~nados-Teitelboim-Zanelli (BTZ)  black hole 
by counting states of the 2d conformal theory living on the boundary
of AdS$_{3}$ \cite {St}. A nice feature of this microscopical 
derivation of the 
black hole entropy is that it does not use string theory or supersymmetry,
but just general properties of 3d gravity. 

Gravity in three spacetime dimensions is rather peculiar, it is a 
topological theory.  For instance, the gauge theoretical formulation of 3d 
gravity
has been used by Carlip \cite{Car1}  
to give a further statistical derivation of the BTZ black hole entropy 
that relies neither on supersymmetry nor on  string theory. It is 
therefore  of interest to study the other low-dimensional element  
of the set  of the AdS/CFT dualities, namely $d=2$, in order to see
if the features of  $d=3$ survive in  this case.

There are also other reasons to study the \adsd/CFT$_{1}$ correspondence.
\adsd appears as solution of a broad class of 2d dilaton gravity 
theories. The same theories have been widely used in the past years to 
investigate
black hole physics in a simplified context \cite{St2}. Moreover, it 
appears as near-horizon description of a variety of black solutions
of string theory. The simplest case, actually, comes from general 
relativity. The near-horizon geometry of the extremal Reissner-Nordstr\"om
black hole,  the Bertotti-Robinson solution, is \adsd$\times S^{2}$.
 
Presently, we do not know very much about the  \adsd/CFT$_{1}$ duality.
Previous work on the subject (and related topics) concerned mainly the asymptotic 
symmetries of \adsd \cite {ho}, string theory on \adsd \cite{st1}, 
or the use of the conformal symmetry to describe the near-horizon 
regime of black holes \cite{so}. 
  
In this paper we present a detailed investigation of the asymptotic 
symmetries of \adsd, their relationship with the conformal group in 
$d=1$ and their use for deriving microscopically the
entropy of 2d black holes. We use as a framework for  our investigation
2d dilaton gravity 
models. Among others, we analyse in detail the  Jackiw-Teitelboim (JT) 
\cite {JT}
model. This choice is motivated by the fact that it is the
simplest 2d gravity model that admits \adsd as solution. 
Moreover, in the context of the JT  model,
\adsd  can be thought of as a $S^{1}$ compactification of AdS$_{3}$,
with the dilaton playing the role of the radius of $S^{1}$.
Some of the results presented here have been  anticipated in a 
previous paper \cite{CM1}.

The structure of the paper is as follows. In Sect. 2 
we discuss the asymptotic symmetries of \adsd. In particular we show 
how the SL(2,R) isometry group of \adsd can be promoted to an infinite
dimensional symmetry on the boundary, generated by a Virasoro algebra.
In Sect. 3 we discuss the conformal group in $d=1$ and its 
relationship with the asymptotic symmetries of \adsd .
In Sect. 4 we discuss the canonical realization  of the symmetries 
and calculate the central charge of the Virasoro algebra.
 In Sect. 5 we give a microscopical 
derivation
of the entropy of the black hole solutions of the JT model and compare 
it with the thermodynamical entropy. In Sect. 6 we extend the 
derivation 
of Sect. 5 to generic black hole solutions that are asymptotically 
Anti-de Sitter. Finally in Sect. 7 we discuss the \adsd/CFT$_{1}$
duality.
   
\section{ Asymptotic symmetries of AdS$_{2}$ }
\paragraph{}

Two-dimensional spacetimes that are asymptotically Anti-de Sitter 
appear as  dynamical solutions of dilatonic gravity  in two 
dimensions \cite{CM,CM2}. Among these  of particular interest are  spacetimes 
with constant negative  curvature (in the following they will be 
referred to as \adsd) 
\be\lb{e1}
R=-2\lq.
\ee
Anti-de Sitter spaces are, for instance, solutions of the JT model, 
whose action is
\be\lb{e2}
A={1\over2}\int \sg \, d^2x\, \e\left(R+2\lq\right),    
\ee
where $\e$ is a scalar
field related to the usual definition of the dilaton $\phi$ by 
$\e=\exp(-2\phi)$.
The geometrical and topological properties of \adsd have 
been already discussed in the literature \cite{CM,CM2}. Here we will just briefly   
remind to the reader  those features that are relevant for our 
discussion. Owing to Birkhoff's theorem of 2d dilaton gravity  the 
general solution of Eq. (\ref {e1}), in a Schwarzschild 
gauge, takes  the form 
 
\be\lb{e3}
\ds-(\l^2x^2-a^2)dt^2+(\l^2x^2-a^2)^{-1}dx^2,
\ee
while the dilaton is given by
\be\lb{e5}
\qquad \e=\eo \l x,
\ee
where $\eo$ and  $a^2$  are integration constants.
Two-dimensional dilaton gravity does not allow a dimensionful analog 
of the Newton constant. However, it is evident from the action 
(\ref{e2}) that the inverse of the scalar field $\e$ represents the
(coordinate 
dependent) coupling constant of the theory, whereas the inverse of
the integration constant
$\eo$  in Eq. (\ref{e5}) plays the role of a dimensionless 2d Newton constant.
 
All the solutions (\ref{e3}) are locally \ads, but have different global
properties. They represent different 
parametrizations of the same manifold, with coordinates patches covering 
different regions of the space. In fact, given the generic solution, 
one can always find a coordinate transformation that 
brings the metric into the form \cite{CM}
\be\lb{e4}
\ds-(\l^2x^2+1)dt^2+(\l^2x^2+1)^{-1}dx^2,
\ee
which represents  full \adsd, a nonsingular, geodesically complete 
spacetime.
At first sight the equivalence of all the metrics (\ref{e3}) up to 
coordinate transformations seems to indicate that the solution to Eq.
(\ref{e1}) is unique and makes it very difficult to interpret the 
solutions
with $a^{2}>0$ as 2d black holes. 
On the other hand, in the context of 2d dilaton  gravity models, 
also the scalar $\e$ must be taken into account
in the discussion of the causal structure of the spacetime.
In general a non-constant  dilaton represents an obstruction that prevents
a maximal 
extension of the spacetime. 
If one considers for instance the solutions (\ref{e3}), (\ref{e5}) 
of the JT model,
one sees that
positivity of $\e$ prevents the analytical continuation of the spacetime 
beyond $x=0$. Moreover, for $x=0$ the (coordinate-dependent) coupling 
constant of the theory diverges, therefore $x=0$  has to be considered 
as a ``singularity'' of the 2d spacetime.

For this reason, in the context of 2d dilaton gravity, one must
regard the solutions with $a^{2}$ positive, negative or zero as 
physically nonequivalent. Following the notation of Ref. \cite{CM}
the corresponding spacetimes will be respectively denoted  by 
\adsdp,\adsdm, \adsdz. 
\adsdp can be interpreted as a black hole with a $x=0$ ``singularity'', 
a $x=\infty$ timelike boundary, an event horizon at $x= a/\l $ and 
a mass given by
\be\lb{mass}
M={1\over 2} \eo a^2\l.
\ee

\adsdz  can be considered as the ground state, zero mass 
solution and the $x=0$ ''singularity'' becomes lightlike. Finally,  
for \adsdm,  $x=0$ becomes  timelike.
   
It is important to stress that owing to the presence of the
dilaton  the global topology of both
\adsdz and \adsdp is different from that of full AdS$_{2}$.
The metric (\ref {e4}) represents a geodesically complete
spacetime of cylindrical topology with two timelike boundaries,
whereas both  \adsdp and \adsdz, due to the 
$x=0$ singularity,  have to be considered as singular
spacetimes with only one timelike boundary at $x=\infty$.

\adsd is a maximally symmetric space, it admits, therefore, three Killing 
vectors generating the $SO(1,2)\sim SL(2,R)$ group of isometries.
In the case of \adsdz  the three
Killing vectors have the form
\be\lb{e6}
^{(1)}\x={1\over \l}{\de\over\de t},\qquad\qquad^{(2)}\x=t{\de\over\de t}-
x{\de\over\de x},\qquad\qquad^{(3)}\x=\l \left(t^2+{1\over\l^4x^2}
\right){\de\over\de t}-2\l tx{\de\over\de x}.
\ee
In the case of \adsdp and \adsdm the $SL(2,R)$ symmetry is 
realised differently than in Eq. (\ref{e6}). However, all the solutions
(\ref{e3}) admit the Killing vector ${\partial \over \l \partial t}$. 

The asymptotic symmetries of \adsd  are by definition the subgroup 
of the 2d diffeomorphisms group that leaves the metric asymptotically 
invariant. Actually, because we are considering \adsd as dynamical solutions 
of 2d dilaton gravity, we should impose asymptotic invariance also on 
the scalar $\e$. This condition for a scalar is in general too 
restrictive. In the following we will therefore use  a milder 
condition on the behaviour of $\e$ under the transformations of the
 asymptotic symmetry group.
A useful application of the concept of asymptotic symmetry is to use 
it  to define the ``global'' charges of 
the   theory. The natural framework for the definition of the charges 
is the Hamiltonian formalism, where they appear as generators of the 
asymptotic symmetries. The canonical realization of the asymptotic 
symmetries will be discussed in  Sect. 4.

By requiring the metric to be of  the form (\ref {e3}), one finds that 
the asymptotic symmetry group of \adsd is the group of 
time-translations ${\cal T}$  generated by the Killing vector $^{(1)}\x$ 
in
Eq. (\ref{e6}). The global charge associated with this symmetry is just the
Arnowitt-Deser-Misner (ADM) mass of the solution (\ref{mass}). 
This requirement appears however too restrictive and does not 
correspond to the intuitive notion of ``asymptotically Anti-de Sitter''. 
As we shall see in Sect. 6, there are solutions of 2d dilaton gravity
that describe spaces whose curvature is only asymptotically constant 
and  differs from Eq. (\ref{e1}) by terms of $o(x^\mo)$. 
One has to choose boundary conditions for the metric at $x\to \infty$
 such that 
 they correspond to the intuitive notion of ``asymptotically
Anti-de Sitter'', are weak enough to enlarge the asymptotic symmetry 
group to a group larger than ${\cal T}$ but tight enough to allow for a 
well-defined definition of the charges associated with the symmetry.
The previous requirements single out the following boundary conditions
on the metric for $x\to \infty$,
\be\lb{e7}
g_{tt}\sim -\l^2x^2+o(1),\qquad g_{tx}\sim\ord{1\over x^{3}},\qquad
g_{xx}\sim{1\over\l^2x^2}+\ord{1\over x^4}.
\ee  
Solving the Killing equations for metrics of the form (\ref {e7}), one 
finds that the asymptotic symmetries are generated by the Killing 
vectors
\be\lb{e8}
\x^t=T(t)+{1\over 2 \l^{4}}{d^{2} T(t)\over dt^2} {1\over x^{2}}+
\ord{1\over x^4},
\qquad\qquad\x^x=- {dT(t)\over dt}\, x+\ord{1\over x},
\ee 
where $T$ is an arbitrary function of the time $t$.

The asymptotic symmetries of \adsd appear as the subgroup of the diffeomorphisms 
of the ``bulk' 2d gravity theory, defined by Eq. (\ref {e8}).
It is evident from the expression of the Killing vectors that these 
symmetries act in a natural way on the one-dimensional, timelike, 
$x\to \infty$, boundary of \adsd as time-reparametrizations. This point 
will be discussed in detail in Sect. 3.
Notice that diffeomorphisms of the ``bulk''  with $T=0$ represent "pure"
gauge transformations on the boundary, because they 
fall off rapidly as $x\to \i$. 

The boundary conditions (\ref{e7}) must be completed by giving boundary 
conditions for the dilaton. In general  the solution for the 
dilaton is model-dependent, here we consider only the case of a 
linear scalar field $\e$ of the form (\ref{e5}). This is the most 
common  case in the context of 2d dilaton gravity models.
The variation of the scalar field $\y$ under the transformations 
(\ref{e8}) is given by $\lie_\x\y=\x^\m\de_\m\y$,
which is asymptotically $o(x)$ for $\y$ of the form (\ref{e5}), and hence of 
the same
order as the field itself. This is quite disturbing, but is an inescapable
consequence of the scalar nature of the dilaton. 
Thus, we require the asymptotic behaviour of the dilaton to be
\be\lb{e9}
\y\sim o(x).
\ee      

The appearance of the  function $T(t)$ in Eq. (\ref {e8}) 
indicates that the asymptotic symmetry group of \adsd is generated by 
an infinite dimensional algebra.  
Since  \ads space has a natural periodicity in $t$, it is convenient to
expand the function $T(t)$ in a Fourier series in the interval 
$0<t<2\p/\l$.
The generators of the asymptotic symmetries then read,
\bea\lb{e10}
A_k&=&\ul\left[1-{k^2\over2\l^2x^2}+\ord{1\over x^4}\right]\cos(k\l t)
{\de\over\de t}+\left[kx+\ord{1\over x}\right]\sin(k\l t){\de\over\de x},
\nonumber\\
B_k&=&\ul\left[1-{k^2\over2\l^2x^2}+\ord{1\over x^4}\right]\sin(k\l t)
{\de\over\de t}-\left[kx+\ord{1\over x}\right]\cos(k\l t){\de\over\de x},
\eea
where $k$ is an integer. One can easily verify that the generators 
satisfy the commutation relations,
\bea\lb{e11}
& &[A_k,A_l]=\ha(k-l)B_{k+l}+\ha(k+l)B_{k-l},\nonumber\\
& &[B_k,B_l]=-\ha(k-l)B_{k+l}+\ha(k+l)B_{k-l},\nonumber\\
& &[A_k,B_l]=-\ha(k-l)A_{k+l}+\ha(k+l)A_{k-l}.
\eea
The algebra can be put in a more familiar form by   defining new generators 
$L_k=-(B_k-iA_k)$, 

\be\lb{e12}
[L_k,L_l]=(k-l)L_{k+l}.
\ee   
The algebra (\ref{e12}) is easily recognised as a Virasoro algebra.
As expected, it contains as a subalgebra the $SL(2,R)$ algebra,
generated by $L_{0},L_{1}, L_{-1}$.

It is interesting to notice that the algebra generating the asymptotic 
symmetries of \adsd can be thought of as ``half'' of the algebra 
generating the asymptotic symmetries of \adst. In fact the asymptotic 
symmetries of \adst are generated by two copies of the Virasoro algebra
\cite{Bh}. This is exactly the symmetry of the 2d conformal theory 
living on the 2d boundary of \adst. As we shall see in the next 
sections this fact plays an important role in the discussion of 
the theory living on the one-dimensional boundary of \adsd that should 
give a realization of the symmetry (\ref {e12}). 

The interpretation of the asymptotic symmetries of \adsd as half of the 
symmetries of a 2d conformal field theory  has a natural explanation 
using light-cone coordinates to parametrize \adsd. 
In the conformal gauge 
\be\lb{e13}
ds^2=-e^{2\rho}d\xn d\xm,
\ee
the metric of \adsd can be written in the form \cite{CM,CM2}
\be\lb{e14}
e^{2\rho}={4\over\l^2}(\xn-\xm)^{-2}.
\ee 

In these coordinates the $x\to \infty$ boundary is defined by $\xn =
\xm$. By translating the boundary conditions (\ref{e7}) and solving the 
Killing equations in the new coordinates, one finds that the 
asymptotic symmetries are generated 
by the Killing vectors
\be\lb{e15}
\x^+=T^{+}(\xn)+
\ord{\xn-\xm},
\qquad\qquad \x^-=T^{-}(\xm)+
\ord{\xn-\xm},
\ee
where $T^{+}=T^{-}$ is an arbitrary function of $\xn=\xm$.
When  $T^{+}$ and $T^{-}$ are two independent functions 
of the two light-cone coordinates (in 2d Euclidean space the
holomorphic and antiholomorphic coordinates), Eq. (\ref {e15}) defines 
the conformal group in 2d. 
The action of the conformal group in 2d factorizes into 
independent actions on $\xm$ and $\xn$. It follows that the asymptotic 
symmetries (\ref{e15}) can be considered as the subgroup of the 
conformal group in 2d acting on the (timelike) curve $\xm=\xn$ and 
characterised by the same action on holomorphic and antiholomorphic 
coordinates.   

\section{ Conformal group in d=1 }
\paragraph{}

The conformal group on a flat one-dimensional (1d), timelike, background 
is usually  defined
\cite{DFF} as the group $SO(1,2)\sim SL(2,R)$, generated by the three 
generators $H$, $D$ and $K$, which satisfy the algebra
\be\lb{f1}
[H,D]=H,\qquad\qquad[K,D]=-K,\qquad\qquad[H,K]=2d.
\ee
The generator $H$ acts as time translation $t\to t+a$, $D$ as time dilatation 
$t\to bt$ and $K$ as a combination of translations and a time inversion 
$t\to c/t$. They can be realized as differential operators acting on
1d flat space by writing 
\be\lb{f2}
H={\de\over\de t},\qquad\qquad D=t{\de\over\de t},\qquad\qquad     
K=t^2{\de\over\de t}.\qquad\qquad   
\ee
The generic transformation can also be written in a $SL(2,R)$  
fractional form as
\be\lb{f3}
t'={\a t+\b\over\g t+\d}\qquad{\rm with}\ \ \a\d-\b\g=1,
\ee
where $\a$, $\b$, $\g$, $\d$ are real parameters.

The lagrangian 
\be\lb{f4}
L=\ha\left(\dot\f^2-{g\over\f^2}\right),
\ee
where $g$ is a dimensionless coupling constant,
is invariant under the 1d conformal group and can be easily quantized.
The quantum version of (\ref{f4}) has been largely studied under the name of 
conformal quantum mechanics \cite{DFF,Fub}.

It is interesting to see how the notion of conformal invariance generalizes 
when passing to a general covariant setting. In curved spacetimes, conformal 
invariance is defined through the existence of conformal Killing vectors 
$\x_\m$ such that
\be\lb{f5}
\na_\m\x_\n+\na_\n\x_\m=\a(x^\m)\gmn,
\ee
where $\a(x^\m)$ is an arbitrary function of the spacetime coordinates. It is
evident that in one dimension this condition is empty, and hence $\x_t=\x$
can be any
function of $t$, namely, conformal invariance coincides with invariance under 
diffeomorphisms.
In particular, one may expand $\x=f(t)$ in Laurent series\footnote
{Alternatively, if $t$ is periodic, one can of course expand in Fourier series.},
so that the
Killing vectors $\x^{(k)}=t^{k+1}\de/\de t$ satisfy the Virasoro algebra
\be\lb{f6}
[\x^{(k)},\x^{(l)}]=(l-k)\x^{(k+l)}.
\ee
One easily sees that the subalgebra generated by $L_{-1}$, $L_0$ and $L_1$
coincides with (\ref{f1}).

From the previous discussion it follows that this symmetry is realized by
any generally covariant theory in one dimension. Let us for instance consider 
a scalar field in 1d "curved" space. Its lagrangian can be written as
\be\lb{f7}
L={1\over2e}\dot\f^2+\ha eV(\f),
\ee
where $e$  is the 1-bein. For $V(\f)=m^2$, this can be interpreted as the lagrangian 
for a free particle moving in a higher dimensional curved spacetime \cite{GSW}. 
The \eom give 
\be\lb{f8} 
\left(\dot\f\over e\right)^2=V(\f),\qquad\qquad
{d\over dt}\left({\dot\f\over e}\right)={e\over2}{dV(\f)\over d\f}.
\ee
Under the transformations (\ref{f6}), $\d e=\xk\dot e+\dot\x^{(k)}e$, 
$\d\f=\xk\dot\f$, and
the lagrangian changes by a total derivative
\be
\d L=\ha{d\over dt}\left({\xk\dot\f^2\over e}+\xk eV\right)={d\L\over dt}.
\ee
As usual, it is possible to associate to these symmetries the conserved
currents $D^{(k)}$:
\be
D^{(k)}={\de L\over\de\dot\f}\d\f-\L=
{e\over2}\xk\left({\dot\f^2\over e^2}-V(\f)\right).
\ee
It is easy to see that all the currents vanish due to the \eom (\ref{f8})
and the symmetry is trivially realized.

If one tries to quantize the model, one should fix the gauge and this
of course spoils the invariance. Presently,  we are not aware of any 
quantum mechanical
model that realizes the symmetries (\ref{f6}) in a non-trivial fashion.

\section{ Canonical realization of the asymptotic symmetries }
\paragraph{}
The connection between asymptotic symmetries and global charges of a theory is
well-known \cite{Regge}. In particular, in the hamiltonian formalism the
global charges
appear as generators of the \as of the theory. In order to discuss 
the implications for 2d Anti-de Sitter gravity,
we briefly recall the hamiltonian formulation of the JT model \cite{Kuchar}.
With the parametrization
\be\lb{g0}
\ds-N^2dt^2+\s^2(dx+N^xdt)^2,
\ee
the hamiltonian of the JT theory reads 
\be\lb{g1}
H=\int dx(N\Ht+N^x\Hx).
\ee
$N$ and $N^x$ act, as usual, as Lagrange multipliers enforcing the constraints,
\bea\lb{g2}
\Ht&=&-\Pe\Ps+\s^\mo\y''-\s^{-2}\s'\y'-\l^2\s\y=0,\nonumber\\
\Hx&=&\Pe\y'-\s\Ps'=0,
\eea
where
\be\lb{g3}
\Pe=N^\mo(-\dot\s+(N^x\s)'),\qquad\qquad \Ps=N^\mo(-\dot\y+N^x\y'),
\ee
are the momenta canonically conjugate to $\y$ and $\s$, respectively.
A dot denotes derivative with respect to $t$ and a prime with respect to $x$.

When the spacelike slices are non-compact, however, 
in order to have well-defined variational derivatives, one must add
to the hamiltonian a surface term $J$, which in general depends on the
boundary conditions imposed on the fields \cite{Regge}.
In our case, the boundary reduces to a
point and the variation  $\d J$ of the surface term must be given by\footnote
{As we shall discuss in the following, this variation is well defined 
only when an integration on $t$ is performed.}
\be\lb{g4}
\d J=-\lim_{x\to\inf}[N(\s^\mo\d\y'-\s^{-2}\y'\d\s)-N'(\s^\mo\d\y)+
N^x(\Pe\d\y-\s\d\Ps)].
\ee
One can then evaluate the Hamilton equations, which read
\bea
\dot\Ps&=&\l^2\y N-\s^{-2}\y'N'+\Ps'N^x, \nonumber\\
\dot\Pe&=&\l^2\s N-\s^{-2}\s'N'-\s^\mo N''+\Pe'N^x+\Pe{N^x}'
\eea
together with (\ref{g3}).

In the hamiltonian formalism, the symmetries associated with the Killing
vectors $\x^\m$ are generated by the
phase space functionals $H[\x]$, defined as \cite{Teitel}
\be\lb{g5}
H[\x]=\int dx \big(\xo\Ht+\xp\Hx\big)+J[\x],
\ee
where $\xo=N\x^t$, $\xp=\x^x+N^x\x^t$. The surface term $J[\x]$ can be
interpreted as the charge associated with the Killing vector $\x^\m$.

The charges $J[\x]$ are defined up to the addition of an arbitrary constant, 
and this ambiguity signals the possible appearance of central charges in the 
realization of the symmetries. In fact,
the Poisson bracket algebra of $H[\x]$ yields in general a projective 
representation of the asymptotic symmetry group \cite{Bh}:
\be\lb{g7}
\{H[\x],H[\o]\}=H[[\x,\o]] +c(\x,\o),
\ee
where $c(\x,\o)$ are the central charges of the algebra. 

In view of the boundary conditions discussed in Sect.  2, in our case
the functional $J[\x]$ can be written in finite form as
\be\lb{g6}
J[\x]=\lim_{x\to\inf} \eo \left [- (\l x)\xo (\y'-\l)+(\l x)
{\partial \xo\over \partial r}(\y-\l x)+ {\l^{4}x^{3}\over 2}  \xo 
\left ( g_{xx}- {1\over \la 
x^{2}}\right) +{1\over \l x} \xp \Ps \right],
\ee
where the arbitrary constants have been adjusted so that the charges
vanish for \adsdz. With this definition, one has for \adsdp
\be
J[A_0]={a^2\y_0\over2},\qquad J[A_k]={a^2\y_0\over2}\cos(k\l t),
\qquad J[b_k]={a^2\y_0\over2}\sin(k\l t).
\ee
Hence, $J[A_0]$ equals $M/\l$, where $M$ is the ADM mass of the \adsdp
black hole,
while the other charges are time-dependent. We shall comment on this in a 
moment.

We still have to evaluate the central charge. The calculation can be
performed in two different ways:
one can either compute explicitly the Poisson brackets (\ref{g7}), or  can fix
the gauge so that the constraints $\Ht=0$, $\Hx=0$ hold strongly. In the latter 
case, the charges $J[\x]$ give themselves
a realization of the asymptotic symmetry group through  the
Dirac brackets, namely \cite{Bh},
\be\lb{g8}
\{J[\x],J[\o]\}_{DB}=J[[\x,\o]] +c(\x,\o).
\ee
But the Dirac brackets can also be expressed in terms of the variation of $J[\x]$
under surface deformations as
\be\lb{g9}
\delta_\o J[\x]=\{J[\x],J[\o]\}_{DB}.
\ee
Comparing (\ref{g8}) and (\ref{g9}) and
evaluating them on a background with vanishing charges, one can then obtain the
central charge $c(\x,\o)$ as the charge $J[\x]$ evaluated on the 
surface deformed by $\o$.

In two dimensions, however, the previous calculation runs into problems.
In fact, being the boundary
a point,  the functional derivatives appearing in the Poisson bracket
(\ref {g7}) can be defined only for pure gauge transformations, for which 
the charge $J[\x]$ vanishes. Moreover, the Dirac brackets (\ref{g8}) have no
meaning as
long as the $x\to \i$ boundary is  a point. As a consequence, the 
surface deformation algebra has no definite action on the charges $J[\x]$,
and Eqs. (\ref{g8}) and (\ref{g9}) cannot be used to calculate the central
charge.

One can remedy these difficulties by defining the time-independent 
charges
\be\lb{g10}
\hat J[\x]={\l\over 2\pi} \int_0^{2\pi/\l} dt\, J[\x].
\ee
The functional derivatives  of $\hat J[\x]$ can be easily  defined,
so that the Dirac bracket algebra $\{\hat J[\x],\hat J[\o]\}_{DB}$ 
has a definite meaning. One can also  verify that the action 
of the surface deformation on the charges $\hat J[\x]$ gives a realization
of the  algebra (\ref{e11}).

Replacing in Eq. (\ref{g8}) and (\ref{g9}) the charges $J[\x]$ with
$\hat J[\x]$,
and evaluating on a \adsdz background, one can easily
calculate the central charges . One gets,
\be\lb{g11}
c(A_k,A_l)=c(B_k,B_l)=0, \qquad c(A_k, B_l)= \eo k^3 \d_{|k|\,|l|}.
\ee

Also a direct calculation of the \pb (\ref{g7}) requires the introduction of a
time integration in order to be well defined when $J[\x]$ does not vanish.
Defining $\hat H$ in analogy with $\hat J$, a straightforward calculation
gives, taking into account the asymptotic conditions imposed on the fields,
\be
\{\hat H[\x],\hat H[\o]\}=\hat H[[\x,\o]]+\lim_{x\to\inf}{\l\over2\p}\int 
_0^{2\pi/\l}dt\ [(\xo'\op-\oo'\xp)\l^2x-(\xo\op-\oo\xp)\l]
\ee
Evaluating the integral in the above expression, one recovers the values
(\ref{g11}) for the central charges.

As noticed before,
apart from $J[A_0]$, which gives the mass $M$ of the solution, the other
charges $J[A_k]$ are in general time-dependent. This means that besides 
the mass there are  no conserved quantities. This fact is strongly related to
the presence of the dilaton and its behaviour under the 
transformations (\ref {e10}). On the other hand all the charges $\hat J$ 
 vanish when evaluated on \adsdp  with the  exception of  $\hat 
 J[A_0]$.
  They represent a sort of time-averaged 
charges that can be used to give a canonical representation
of the  algebra (\ref {e11}).

Finally, we notice that
defining the new generators $L_k=-(B_k-iA_k)$, and shifting $L_0$ by  a 
constant, $L_0\to L_0-\y_0$, one obtains the standard form of the Virasoro
algebra with central charge,
\be\lb{g12}
[L_k,L_l]=(k-l)L_{k+l} + {c\over 12}(k^3-k)\d_{k+l},\quad c=24\e_0.
\ee

\section{ Statistical entropy of 2d black holes }
\paragraph{}

A nice application of the results of the previous sections is a
microscopical computation of the entropy of 2d black holes. Before 
calculating the entropy microscopically, let us  first review some 
general facts and the peculiarity of the thermodynamical entropy in 
two-dimensions. In two spacetime dimensions we do not have an area law 
for the black hole entropy. The thermodynamical  entropy 
can be computed using several methods. For instance, using Noether 
charge techniques one finds the general formula \cite{My}
\be\lb {h1}
S=-2\pi Y^{\mu \nu \rho\sigma}\epsilon _{\mu\nu}\epsilon 
_{\rho\sigma}|_{h},
\ee
where the subscript $h$ means that the expression has to be evaluated 
at the black hole horizon and the tensor $Y^{\mu \nu \rho\sigma}$ is defined 
in terms of the 
derivatives of the Lagrangian $L$, characterizing the model, with respect 
to the curvature tensor
$R_{\mu \nu \rho\sigma}$:
\be
Y^{\mu \nu \rho\sigma}= {\partial L\over \partial R_{\mu \nu \rho\sigma}}.
\ee
Using Eq. (\ref{h1}) one finds for the entropy associated with the 
black solutions of a generic 2d dilaton gravity model,
\be\lb {h2}
S=2\pi {\e_h},
\ee
where $\e_h$ is the value of the scalar field $\e$  at the horizon.
Although in two spacetime dimensions we do not have an area law for 
the black hole entropy, Eq. (\ref{h2}) can be interpreted as
a generalization to 2d of the Bekenstein-Hawking entropy.
This follows simply from the fact that  according to Eq. (\ref{e5}),
$\e$ is nothing but the "radial" coordinate of the 2d space.

Using Eq. (\ref{h2}) one can easily calculate the thermodynamical 
entropy
associated with the black hole  (\ref{e3}) .
We have \cite{CM}:
\be\lb{h3}
S=4\pi \sqrt {\eo M\over 2\l}.
\ee
One can also derive the same result by integrating the thermodynamical 
relation $T dS=dM$, given the temperature $T$ as a function of the 
mass $M$.
 
Our derivation of the statistical entropy of 2d black holes follows 
closely that of Strominger for the BTZ black hole \cite {St}. We just need 
to count the 
excitations around the vacuum \adsdz with $M$ given, in the 
semiclassical approximation of large $M$.
Because states on the 2d bulk are in correspondence with states 
living on 
the $x\to \infty $, boundary we can equivalently count excitations on the 
boundary. In the semiclassical regime $M/\l>>c$, 
the density of states  is given by the Cardy's formula \cite{Ca}:       
\be\lb{h4}
S=2\pi \sqrt{c\,l_0\over 6},
\ee
where $l_0$ is the eigenvalue of the Virasoro generator $L_0$, which
for a black hole of mass $M$ is given by\footnote
{The shift in $L_0$ performed in the previous section in order to obtain
the Virasoro algebra in its standard form can be neglected for large $M$.}

\be\lb{h5}
l_0= {M\over\l}.
\ee
Inserting Eq. (\ref{h5}) and the value of the central charge $c$ given 
by Eq.
(\ref{g12}) into Eq. (\ref {h4}), we find for the statistical entropy, 

\be\lb{h5a}
S= 4\pi \sqrt {\eo M\over \l},
\ee
which agrees, up to a factor $\sqrt 2$,  with the thermodynamical 
result (\ref{h3}). 

Because we do not have an explicit representation of the degrees of 
freedom living in the boundary, it is very   
difficult to  explain  the discrepancy 
between the statistical  and the thermodynamical result. Nevertheless,  
a simple justification
of the factor  $\sqrt{2}$ can be found  if one considers 
the model (\ref{e2}) as a circular symmetric dimensional reduction of 
three-dimensional 
gravity, with  the field $\e$ parametrizing the radius of the circle.
Using the notation of Ref. \cite{St}, the 2d dilaton gravity action (\ref{e2})
can be obtained from the 3d one by means of the ansatz,
\be\lb{h6}
ds^{2}_{(3)}= ds^{2}_{(2)}+ 16G\e^{2}d\varphi^{2},
\ee
where $G$ is the 3d Newton constant and $0\le\varphi\le 
2\pi$.
The 2d black hole (\ref{e3}) can be 
considered as the dimensional reduction of the zero angular momentum 
($J=0$) 
BTZ black hole. Simple calculations show that both the mass and the 
thermodynamical entropy of the BTZ black hole agree with our 2d 
results, 
given respectively by Eq. (\ref{mass}) and  Eq. (\ref{h3}).
The same is not true for the statistical entropy. From 
the 3d point of view we have contributions to the mass of the 2d black 
hole coming from both the right- and left-movers oscillators of 
the 2d conformal field theory living on the boundary of AdS$_{3}$. 
Since $J=0$ implies that the number of right-movers equals that of 
left-movers, we have $l_{0}=M/2\l$, which inserted into the Cardy's 
formula reproduces the thermodynamical entropy (\ref{h3}).
From the 2d point of view only oscillators of one sector contribute 
to the mass of the black hole giving $l_{0}=M/\l$ and the statistical 
entropy (\ref{h5a}). 

These results are in accordance with those obtained 
by Strominger in Ref.  \cite {st1}, where AdS$_{2}$ is 
generated as the near-horizon, near-extremal limit  of AdS$_3$.
At first sight  these results seem to imply that there is no intrinsical 2d 
explanation of the statistical entropy of 2d black holes.
This is certainly true as long as  the field
$\e$ is interpreted as the radius of the internal circle, because the 
$x\to\infty$ boundary of AdS$_2$ corresponds to the $\e\to \infty$  region, 
where   
the internal circle decompactifies and the 2d theory becomes intrinsically 3d. 

The previous considerations do not apply when AdS$_2$ arises as 
near-horizon geometry of higher dimensional black holes with no 
intermediate AdS$_3$  geometry  involved. We do not have a complete
explanation of the factor  $\sqrt 2$ in this case. In our opinion what
one needs in order to find an explanation of this discrepancy 
is a complete understanding of the role played in our derivation by the 
global topology of AdS$_2$. Full AdS$_2$ has a cylindrical topology 
with two  disconnected timelike boundaries. This fact plays a crucial 
role in Ref. \cite{st1} because it  makes the string theory living 
 on AdS$_{2}$
a theory of open strings. By studying the black hole solutions of
the JT theory we are forced to cut the spacetime on the $x=0$ 
``singularity'', so that only one timelike boundary of  full AdS$_{2}$
is available. It seems to us  that a thorough understanding of the 
statistical entropy of 2d black holes will be at hand only when 
this point will be fully clarified. 

Perhaps an answer to this question can be found analyzing 
 2d dilaton 
gravity models that admit AdS solutions with a constant dilaton. 
In this case there is no ``singularity'' and the spacetime  can be
extended to full AdS$_{2}$. Moreover, this is the most interesting case
from the string theoretical point of view, because the near-horizon 
geometry of most of the higher dimensional extremal black hole solution appearing in 
string theory behaves as AdS$_{2}\times $ C, where C is some compact 
manifold and is characterised by a constant dilaton. This case presents some additional
difficulties if compared with that analysed in this paper. A constant 
dilaton makes a black hole interpretation of the solutions very difficult,
at least from the 2d point of view.  

\section {General Models}
\paragraph{}
Until now our considerations have been restricted to the JT model
(\ref{e2}). However our results can be easily extended to more general 
2d dilaton gravity  models that admit solutions satisfying the 
boundary conditions (\ref{e7}).
Let us consider the generic action 
\be\lb{l1}
A={1\over2}\int\sg\ d^2x\ \left[\e R+\lq V(\e)\right],   
\ee
where $V(\e)$ is the dilaton potential. 
The general solutions of the model are \cite{Cadoni}
\be\lb{l2}
\ds-\e_{0}^{-2}\left(J - {2M\e_{0}\over \l}\right )dt^2+
\e_{0}^{2}\left (J - {2M\e_{0}\over \l}\right )^{-1}dx^2,
\qquad \e=\e_{0}\l x,
\ee
where $J=\int V d\e$ and $M$ is the mass of the solution. Under suitable 
conditions the solutions 
(\ref{l2}) can be interpret  as 2d black holes. 
We are interested in black hole solutions that are asymptotically AdS. 
More precisely, the solutions must behave asymptotically as in Eq. 
(\ref {e7}). A simple calculations shows that a sufficient condition for 
this to happen is that the potential $V$ behaves for $\e\to \infty$ as
\be\lb{l3}
V= 2\e + \ord {\e^{{-2}}}.
\ee
The calculations  of the previous sections can be easily generalized 
to  the  class of models whose potential has the asymptotic behavior
(\ref{l3}). Because   the asymptotic symmetries depend only on the 
asymptotic behavior of the metric and of $\e$, it turns out that they 
are exactly the same as those described by Eq. (\ref{e8}).
Moreover, the models (\ref{l1}) differ from the JT model 
(\ref{e2}) only  in the form  of the potential $V$.
Because $V$ does not enter in the calculation of the central charge 
$c$ described in Sect. 4, it follows that also in this general case $c$ 
is given by Eq. (\ref{g12}). Hence, in the semiclassical regime of large $M$ we 
find  that the statistical entropy of the black hole solutions 
(\ref{l2}) is given by Eq. (\ref{h5a}). To compare it with the 
thermodynamical entropy we use Eq. (\ref{h2}). For large $M$ the 
thermodynamical entropy  has the form
\be\lb{l4}
S= 4\pi \sqrt{M\over 2 \l} +\ord{M^{-1}}.
\ee
The leading term in the expansion exactly matches the result (\ref{h3}) 
for the JT black hole.  
 
\section{ The AdS$_{2}$/CFT$_{1}$ correspondence}
\paragraph{}
The meaning of the AdS/CFT duality in two spacetime dimension is
a rather controversial subject. General arguments suggest that 
all 2d gravity theories are conformal field theories. This is not only 
true for quantum theories of gravity  \cite{Po} but there is also some
evidence that this could also be true for classical dilaton gravity \cite{marco}.
Strominger argued that quantum gravity  on  AdS$_{2}$ is described  by a 
Liouville theory, which, owing to the cylindrical topology of the 
space, is essentially an open string theory on the strip \cite{st1}. As a
consequence the $SL(2,R)$ isometry group of   AdS$_{2}$ is enhanced to 
(half of) the 2d conformal group. This is essentially the same 
result we have obtained in the previous sections by analyzing the 
asymptotic symmetries of AdS$_{2}$.

These general arguments, together with those presented in Sect. 5,
suggest  that in the family of the AdS$_{d}$/CFT$_{d-1}$ dualities, 
the $d=2$ case is very similar to the $d=3$ one,  the 
conformal group being in both instances infinite dimensional.
The same general arguments suggest that the CFT$_{1}$ appearing in the  
\adsd/CFT$_{1}$ correspondence 
is some kind of conformal quantum mechanics living in the boundary(ies) 
of \adsd.

In spite of these similarities with the $d=3$ case the \adsd/CFT$_{1}$ correspondence
remains still mysterious and puzzling. One problem is  that 
full \adsd  has two timelike boundaries. The space where the dual conformal theory 
should live is not a connected manifold. By considering \adsdz instead 
of full \adsd we have only  a timelike boundary but we pay the price 
of having a singular, not geodesically complete, spacetime. As we have 
argued in Sect. 5, this is probably the feature that is responsible for
the mismatch between thermodynamical and statistical 
entropy of 2d black holes.
A second feature that is peculiar to the $d=2$ case is the complete equivalence 
of the diffeomorphisms and the conformal group in  one dimension. The 
physical implication of this equivalence is that the usual difference 
between gauge symmetries and symmetries related to conserved 
charges disappears. 
For this reason, as pointed out is Sect. 3,  the task of finding physical systems that 
realize the conformal symmetry becomes very hard to solve.   

There is a simple way in which one may avoid the previous problems.
One just needs to
assume that the $d=2$ case is not fundamental, but ``intrinsically''  
three-dimensional. If one accepts this point of view the CFT$_{1}$ 
should be thought of just as (half) of CFT$_{2}$, in the way  we have explained 
in Sect. 2.  Evidence that this could be true has been given in Ref.
\cite{st1}, where it has been argued that the in the $d=2$ context the 
general AdS$_{d}$/CFT$_{d-1}$ duality becomes a duality between two
2d conformal field theories. 
This  is certainly true when AdS$_{2}$ arises as a $S^{1}$-compactification  
of AdS$_3$ (this is the case analysed in Ref. \cite {st1}). In our 
approach this implies the identification of the field $\e$ with the 
radius of $S^{1}$ (see Sect. 5). Thus, one has  a simple explanation 
of the ``intrinsical'' two-dimensional nature  of the conformal 
symmetry: the $x\to \infty$ boundary corresponds to the spacetime
region where  the  $S^{1}$ decompactifies.
 
However, when \adsd is not generated as compactification of  AdS$_{3}$
(this the case of the near-horizon geometry of a variety of higher
dimensional black holes in string theory) there is no reason why 
AdS$_{3}$ should be more fundamental than AdS$_{2}$ and no reason to 
consider  CFT$_{1}$ as half of CFT$_{2}$.
Furthermore, there are some indication that the theory living in the 
boundary(ies) of AdS$_{2}$ could be a genuine (though somehow exotic)
quantum mechanical theory. The first indications comes from dilatonic 
quantum gravity studies. It has been shown that owing to the so-called
Quantum Birkhoff Theorem 2d dilatonic quantum gravity reduces to 
quantum mechanics \cite{ma}. The quantum mechanical theory arising from 
 dilaton gravity models admitting AdS$_{2}$ as solution, could be a good 
candidate for 
CFT$_{1}$. 
A second indication comes from a recent work of Gibbons and Townsend 
\cite{GT}, where it is argued that the large $n$ limit of $n$-particle 
superconformal Calogero model gives a microscopical description of the
near-horizon limit of extreme Reissner-Nordstr\"om black holes.


\begin{thebibliography}{1}
\bibitem {Su} 
L. Susskind, J. Math. Phys.  {\bf 36}, 6377 (1995).


\bibitem {Wm}
J. Maldacena, Adv. Theor. Math. Phys. {\bf 2}, 231 (1998);
E. Witten, Adv. Theor. Math. Phys. {\bf 2}, 253 (1998);
E. Witten, Adv. Theor. Math. Phys. {\bf 2}, 505 (1998).


\bibitem {Bh} 
J. D. Brown and M. Henneaux, \CMP {\bf 104}, 207 (1986).

\bibitem {St}
A. Strominger, J. High Energy Phys. {\bf 02} (1998) 009.

\bibitem {Car1}
S. Carlip, \PRD {\bf D51}, 632 (1995); \CQG{\bf 15}, 3609 (1998).


\bibitem {St2}
See for instance, A. Strominger in {\sl Les Houches Lectures on
Black holes}, lectures presented at the 1994 Les Houches Summer 
School, hep-th/9501071.


\bibitem {ho}
M. Hotta, ``Asymptotic symmetries and two-dimensional Anti-de Sitter 
gravity'', gr-qc/9809035.

\bibitem{st1}
A. Strominger, ``AdS$_{2}$ Quantum Gravity and String Theory'', 
hep-th/9809027. 

\bibitem{so}
S. Solodukhin, ''Conformal description of horizon states'', 
hep-th/9812056; D. V. Fursaev, `` A Note on Entanglement Entropy and Conformal 
Field Theory'', hep-th/9811122.


\bibitem{JT}
C. Teitelboim, in ``Quantum theory of Gravity'' edited by S.M. 
Christensen (Hilger, Bristol); R. Jackiw, ibid.
 
\bibitem {CM1}
 M. Cadoni, S. Mignemi, ``Entropy of 2d black holes by counting 
 microstates'', hep-th/9810251.

\bibitem {CM}
 M. Cadoni, S. Mignemi \PRD   {\bf D51} (1995) 4319.

\bibitem {CM2}
 M. Cadoni, S. Mignemi \NPB   {\bf B427} (1994) 669.

\bibitem {DFF}
V. De Alfaro, S. Fubini and G. Furlan, \NCA{\bf 34A}, 569 (1976).

\bibitem {Fub}
S. Fubini and E. Rabinovici, \NPB{\bf B245}, 17 (1984);
P. Claus, M. Derix, R. Kallosh, J. Kumar, P.K. Townsend and A. Van Proeyen,
\PRL{\bf 81}, 4553 (1998).

\bibitem {GSW}
M.B. Green, J.H. Schwarz and E. Witten, {\it Superstring theory}, Cambridge
University Press 1987.

\bibitem {Regge}
T. Regge and C. Teitelboim, \AP {\bf 88}, 286 (1974).

\bibitem {Kuchar}
K.V. Kuchar, J.D. Romano and M. Varadarajan, \PRD{\bf D55}, 795 (1997);
D. Louis-Martinez, J. Gegenberg and G. Kunstatter, \PLB{\bf B321}, 193 (1994).


\bibitem{Teitel}
C. Teitelboim, \AP {\bf 79}, 542 (1973). 


\bibitem {My}
See for instance R. C. Myers, \PRD {\bf D50}, 6412 (1994) and 
references therein.



\bibitem {Ca}
J. A. Cardy, \NPB {\bf B270}  (1986) 186.


\bibitem {Cadoni}
See for instance M. Cadoni, \PRD {\bf D53}, 4413 (1996) and 
references therein.

\bibitem {Po}
V. Knizhnik, A. Polyakov, A. Zamolodchikov, \MPLA {\bf A3}, 819 
(1988); F. David  \MPLA {\bf A3}, 1651 (1998).

\bibitem{marco}
M. Cadoni, \PLB {\bf B395}, 10 (1997); M. Cavagli\'a, ``Geometrodynamical 
formulation of Two-dimensional dilaton gravity'', hep-th/9811059.

\bibitem {ma}
M. Cavagli\'a, V. De Alfaro and A. T. Filippov, \PLB {\bf B424}, 265 
(1998).


\bibitem {GT}
G. W. Gibbons and P.K. Townsend, ``Black holes and Calogero models'',
hep-th/9812034.

\end{thebibliography}
\end{document}